\documentstyle[12pt,emlines2,epsfig]{article}
\def\be{\begin{equation}}
\def\ee{\end{equation}}
\newcommand{\zz}{%
\hspace{1mm}\begin{picture}(20,10)%
\put(0,0){\line(1,0){10}}%
\put(4.5,2){\mbox{\scriptsize{/}}}%
\end{picture}\hspace{-2.3mm}}
\newcommand{\cc}{-\circ -}

\begin{document}

\title{Quantum Decoherence and
Higher Order Corrections to the Large Time Exponential
Behaviour}
\author{
I.Ya. Aref'eva\thanks{arefeva@mi.ras.ru} and I.V. Volovich
\thanks{volovich@mi.ras.ru}
\\
{\it Steklov Mathematical
Institute, Russian Academy of Sciences}\\ 
{\it Gubkin St.8, GSP-1, 117966, Moscow, Russia}
}
\date{}
\maketitle

\begin {abstract}
There exists the well known approximate expression describing
the large time behaviour of matrix elements of the evolution
operator in quantum theory: $<U(t)>\simeq\exp(at)$. This expression
plays the crucial role in considerations of  problems
of  quantum decoherence, radiation, decay, scattering theory,
stochastic limit, derivation of master and kinetic equations 
 etc. This expression was obtained in the Weisskopf-Wigner
approximation and 
in the van Hove  (stochastic) limit. We derive the exact general
formula which includes the higher order corrections
to the above approximate 
expression: $<U(t)>=\exp(At+B+C(t))$. The constants
$A$ and $B$ and the oscillating function $C(t)$ are computed
in perturbation theory. The method of perturbation of spectra
and renormalized
wave operators is used. The formula is valid for a general class of
Hamiltonians used in statistical physics and quantum field theory.
\end {abstract}
\newpage
\section{Introduction}
\setcounter{equation}{0}
The study of the large time behaviour of the evolution operator in
statistical physics and quantum field theory is the subject 
of numerous investigations.
The basic object to study in quantum field theory is the scattering matrix.
The physical idea behind the  scattering matrix approach is that 
in the scattering processes there exists a characteristic time scale
such that in a time regime larger  then this time scale one can neglect 
interaction and particles evolve according to the free dynamics \cite{BS}.
However, due to the infinite number of degrees of freedom in quantum field 
theory asymptotic dynamics is not simply governed  by  the free Hamiltonian.
There are effects of renormalization of vacuum energy and one particle
states and the asymptotic
states become in fact the states of dressed particles \cite{SchB}. 
Therefore one has to deal with renormalized, or dressed wave operators
\cite{Frie,LD,Hepp, ARE}.
In this paper we will use the method of renormalized wave operators
to compute matrix elements of the evolution operator for finite time.

There are many important problems in quantum field
theory  where we are interested  in the large but not infinite  time and
 where the standard $S$-matrix description is not 
very convenient or 
even not applicable. These include  processes with unstable  
particles \cite{Sch,GW}
(in fact almost all particles are unstable), atom-photon interactions
\cite{CDG}, elementary particles in "semidressed states" with 
non-equilibrium proper fields \cite{Fei}, electroweak baryogenesis and
phase transitions in the early Universe and in high-energy collisions
\cite{RS}, quantum optics \cite{WM}, 
quantum decoherence (see for example \cite{WM,Zur,Vol1})
 etc. In the consideration
of such processes we are interested in the time regime smaller
 than the "infinite" time when the $S$-matrix description
 becomes applicable.  
The consideration of such processes
belongs to non-equilibrium quantum field theory, see
\cite{AAV} for more discussions.

In statistical physics there is not just one but several relevant 
time scales
and as a result we don't have here a universal method 
comparable with the $S$-matrix approach  in quantum field theory.
One can say that the role of $S$-matix approach in non-equilibrium
statistical physics is played by various master and kinetic
equations.
Various methods of consideration of time evolution for classical and
quantum systems have been developed by Bogoliubov \cite{Bog1},
Weisskopf and Wigner (see \cite{GW}), van Hove \cite{vHo1},
Prigogine \cite{PH} and many others, for a review see for example
\cite{Zub,LP}.

A general method in non-equilibrium statistical physics and 
quantum field theory is
the method of stochastic limit, see \cite{AcKoVo,AcLuVo}.
 The idea of this method is
the systematic application of the $\lambda^2t$-limit and quantum
stochastic differential equations. One 
considers  the evolution operator $U(t)$ of quantum system for 
small coupling constant $\lambda$ and large time $t$.
The limit 
\be
\lambda\to 0,\qquad  t\rightarrow \infty ,\qquad  \lambda^2t
= \mbox {fixed} =\tau
\label{i1}
\ee 
has been considered by Bogoliubov \cite{Bog1},  
Friedrichs \cite{Fri2} and van Hove \cite{vHo1}.
In the quantum theory of open systems, the 
limit (\ref{i1}) is known as the van Hove  or the $\lambda^2t$ limit.

In this paper we study corrections to the (van Hove) stochastic limit.
For this purpose a general formula for the matrix elements of 
the evolution operator is obtained. This formula is used for the 
investigation of the large time and small coupling constant asymptotic 
behaviour. We consider a very general class of  Hamiltonians used in solid 
state physics and quantum field theory
\begin{equation}
H=H_0+\lambda V, \label{1}
\end{equation}
where $H_0$ is a free
Hamiltonian,  $V$ describes an interaction and $\lambda$
is the coupling constant.

It is well known that the large time behaviour of the
vacuum expectation value of the evolution operator 
\begin{equation}
U(t)=e^{itH_0}e^{-itH},\label{2}
\end{equation}
is given by
\begin{equation}
\langle U(t)\rangle\simeq e^{at},
\label{3}
\end{equation}
where $a$ is a constant.
This expression can be obtained in the Weisskopf-Wigner
approximation \cite{GW} or 
in the van Hove  (stochastic) limit.
In this paper we obtain the following exact general formula valid for any
time $t$

\begin{equation}
\langle U(t)\rangle=e^{iAt+B+C(t)}.
\label{I4}
\end{equation}
Here $A$ and $B$ are constants for which a representation
  in perturbation theory will be given and
$C(t)$ is a function which under rather general assumptions 
 can be represented for large time $t$ as
 \begin{equation}
C(t)={f(t)\over t^{\alpha}}.
\label{5}
\end{equation}
Here
$f(t)$ is a bounded oscillating  function
and  the exponent $\alpha$ depends on the model and 
on the dimension of space ($\alpha=3/2$ for the physical
3-dimensional space and for a general class of models).
We derive the main formula (\ref{I4}) by using the theory of perturbation
of  spectra and renormalized wave 
operators.

One of the remarkable features of quantum mechanics which
most distinguishes it from classical mechanics is the coherent
superposition of physical states. The important point is the physical
distinction between the coherent superposition of states and the
classical mixture, see for example \cite{WM,Zur}. In particular
the maintenance of quantum coherence is a crucial requirement
 of the ability of quantum computers to be more efficient
 in certain problems than classical computers (see for example
 \cite{Vol1} and refs therein). Any state of a quantum system
 can be described by the off-diagonal elements of the density
 operator. In fact the dominant contribution for large
 time to any matrix
 element of the evolution operator comes from vacuum
and if $\Re a<0$ in (\ref{3}) then one has the exponential decay.
The constant $a$ is a function of the coupling constant
and other parameters of the model. To suppress decoherence we would like to
have the regime with $\Re a=0$. In such a case we have to investigate
corrections to the approximate expression (\ref{3}). The solution
of this problem is given in this paper and it is presented in (\ref{I4}).

Expectation value in (\ref{I4}) is taken over vacuum. For the case 
of one-particle states we obtain
\be
\langle p| U(t)|p'\rangle=e^{iA(p)t+B(p)+C(t,p)}\delta (p-p').
\label{I4a}
\ee

The formulae (\ref{I4}) and  (\ref{I4a}) have a very general character. 
We prove (\ref{I4}) in 
Section 3 for a very general class of Hamiltonians.
From this formula one gets the stochastic limit of evolution
operator 
\begin{equation}
\langle U(\tau/\lambda^2)\rangle=e^{iA_2\tau}(1+o(\lambda)),
\label{I4s}
\end{equation}
as well as  corrections to
stochastic limit
\be
\langle U(\tau/\lambda^2)\rangle=e^{iA_2\tau+\lambda^2(B_2+i\tau A_4)+
o(\lambda^2)}.
\label{I4n}
\ee

In Section 4 we  calculate 
corrections to the stochastic limit  for the
one particle matrix   elements of the evolution operator
 for the case of translation
invariant Hamiltonians 
\be
\langle p| U(\tau/\lambda^2)|p'\rangle=
e^{iA_2(p)\tau}(1+\lambda^2(B_2(p)+ i\tau A_4(p))+o(\lambda^2))
\delta (p-p')
\label{I4oc}
\ee

The class of considered Hamiltonians includes the Bose and Fermi gases,
phonon self-interaction and electron-phonon interaction, quantum
electrodynamics in external fields etc.

We give two independent derivations of formula (\ref{I4}) and (\ref{I4oc}).
The first method consists in the direct examination of perturbation
theory. The second  one uses powerful results 
of spectral theory and renormalized wave
operators  \cite{ARE}. Although the second method 
is simpler we present both of them since the first one can be used 
also  in the case of decay
when the results of the standard scattering theory are not applicable.

\section{Notations and auxiliary results }
\setcounter{equation}{0}
\subsection{Hamiltonians}

 We consider
Hamiltonians of the  form (\ref{1}) where $H_0$ is a free
Hamiltonian
\be
\label{I1}
H_0=\sum _i \int \omega _i(k)a^*_i(k)a_i(k)d^dk
\ee
and $V$ is the sum of Wick monomials.
Creation and annihilation operators $a^*_i(k)$, $a_i(k)$
describe particles or quasiparticles and they 
satisfy the commutation or anticommutation relations
\be
\label{I2} [a_i(k), a^*_j(k')]_{\pm}=\delta _{ij}\delta (k-k')
\ee
Here $k,k^{'}\in R^d$ and $i,j=1,..,N$ label the finite
number of different types of (quasi)particles.
Examples of one-particle energy $\omega _i(k)$
include the relativistic ($\omega (k)=(k^2+m^2)^{1/2}$)
and non-relativistic ($\omega (k)= k^2/2-\omega_0$) laws, the Bogoliubov
spectrum ($\omega (k)= (bk^4+k^2v(k))^{1/2}$), the Fermi quasiparticle
spectrum ($\omega(k)=|k^2/2m-\mu|$) etc.

 We consider two
different types of Wick polynomials.  The first type describes an
interaction in the case when there is not the  translation invariance

\be
V=\sum_{I,J }
\int v(p_1,i_1\dots p_{I},i_I|q_1,j_1\dots q_J,j_J)
\prod^I_{l=1}a^*_{i_l}(p_l) dp_l\prod^J_{r=1}a_{j_r}(q_r)dq_r \label{4}
\ee
were $v(p_1\dots p_{I}|q_1\dots q_J)$ are some test  functions.

 The second type is described by the translation invariant
Hamiltonian

\be
V=\sum_{I,J}V_{I,J}=\sum_{I,J}\int \hat {v}
(p_1,i_1,\dots p_{I},i_I|q_1,j_1\dots
q_{J-1},j_{J-1},J_j)\label{H8}
\ee

$$
\delta
\left(\sum^I_lp_l-\sum^J_rq_r\right)\prod^I_{l=1}a^*_{i_l}(p_l)
dp_l\prod^J_{r=1}a_{j_r}(q_r)dq_r
$$

Clearly the delta function causes the trouble and there are  singular 
terms in (\ref{H8}). Namely, $V_{I,0}\phi _n $ does not belong to the Fock
space unless $\phi _n=0$.
This singularity is called the volume singularity. To give a meaning
to the Hamiltonian  with interaction (\ref{H8}) 
one has to introduce a volume 
cut-off, then perform the vacuum renormalization and 
vacuum dressing and only after that 
remove the cut-off. This procedure defines the hamiltonian 
in a  new  space (see \cite{Hepp,Are} for details). To avoid this 
difficulty in this paper we  will assume that for translation 
invariant interaction there are no pure creation and annihilation terms.

\subsection{Evolution operator}

We will investigate  the evolution operator
\be
\label{U1a}
U(t)=e^{itH_0}e^{-it(H_0+\lambda V)}.
\ee
 In  perturbation theory the evolution operator (\ref{U1a})
has the representation 
\begin{equation}
U(t)=1-i\lambda\int^t_0V(t_1)dt_1+
(-i\lambda)^2\int^t_0dt_1\int^{t_1}_0dt_2V(t_1)
V(t_2)+\dots
\label{U2}
\end{equation}
where
$$
V(t)=e^{itH_0}Ve^{-itH_0}
$$

We will also use the evolution operator with the adiabatic cut-off
\begin{equation}
U_{\epsilon}(t)=1-i\lambda\int^t_0V_{\epsilon}(t)
U_{\epsilon}(t)dt
\label{U3)}
\end{equation}
$$
V_{\epsilon}(t)=e^{-\epsilon |t|}e^{itH_0}Ve^{-itH_0}$$
In this case  one also has the perturbation series 
\begin{equation}
U_{\epsilon}(t)=1-i\lambda\int^t_0V_{\epsilon}(t)dt+
(-i\lambda)^2\int^t_0dt_1\int^{t_1}_0dt_2V_{\epsilon}(t_1)
V_{\epsilon}(t_2)+\dots .
\label{U4a}
\end{equation}

We will use for $V_{I,J}$ the diagram representation.
The corresponding diagram has one vertex and $I$ lines 
going from the vertex
to the left and $J$ lines going to the right.
The first $I$ lines represent creation operators and the last $J$
lines represent annihilation operators. In what follows we will use
the Wick theorem:
\be
V_{I,J}W_{N,M}=:V_{I,J}W_{N,M}:+\sum _{s}^{\min \{ J,N\}}
V_{I,J} \underbrace{-\circ - }_{s} W_{N,M}.
\label{n10}
\ee
The  kernel of  the Wick monomial  $:V_{I,J}W_{N,M}:$
is
\be
v_{I,J}\otimes w_{N,M}
\label{n11}
\ee
and $V_{I,J}\underbrace{ -\circ - }_{s}W_{N,M}$ is  the Wick monomial

\be
V_{I,J} \underbrace{-\circ -} _{s}W_{N,M}=\int 
\prod _{r=1}^{I+N-s}(dp_ra^*(p_r))
\cdot
\prod _{l=1}^{J+M-s}(dq_la(q_l))C_J^sC_N^s t!\int \prod _{r=1}^{s}dk_r
(v\circ _s w)(k_1,...k_r,p_1,...;q_1...)
\label{N12a}
\ee
with the following   kernel

\be
(v\circ _s w)(k_1,...k_r,p_1,...;q_1...)=
\label{n13}
\ee
$$
v_{I,J}(p_1,...,p_I;k_1,...k_s,q_1,...q_{J-s})
w_{N,M}(k_1,...k_s;p_{I+1},...,p_{I+N-1},q_{J-s+1},...q_{J-s+M}).
$$
Below we will drop out the symbol $s$ in $\circ _s$
and to specify the concrete form of contractions will refer to 
corresponding diagrams.

\begin{figure}[h]
\begin{center}
\epsfig{file=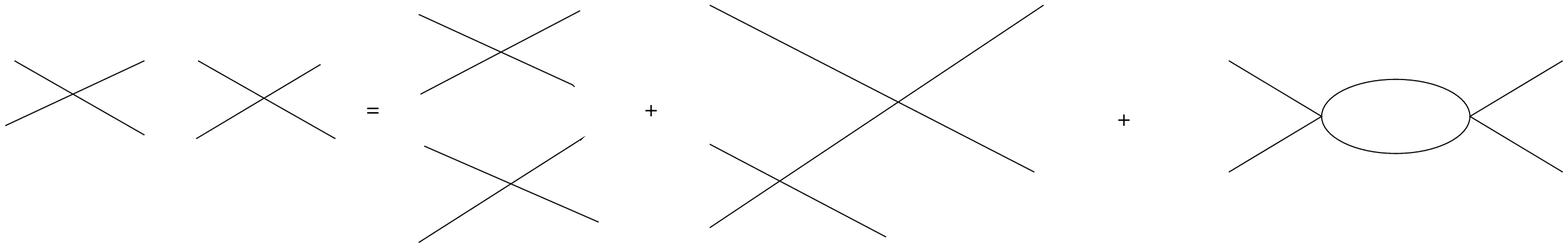,
   width=450pt,
   angle=0
 }
\end{center}
\caption{Graphs representation of (\ref{n10})}
\label{Fign10}
\end{figure}

The equality (\ref{n10}) for the 4-point interaction has the 
diagram representation as shown on
Fig. \ref{Fign10}.

We will use  also the following notations.
The line of the graph is called the internal if it connects two vertices
of the graph.
A graph is the connected graph if 
all its vertices are connected by a set of
internal lines otherwise it is called the  disconnected one.
A connected graph is called the one-particle reducible (1PR)
if after the removing a line it becomes  disconnected.
A connected graph is called the one-particle irreducible (1PI)
if after the removing any line it is still connected.

The following "linked cluster theorem" \cite{Hepp} will be used:

\be
\label{1a}
U(t)=:e^{U_c(t)}\ :
\ee
where
$$
U_c(t)=
\sum_{n=1}^{\infty}
(-i\lambda)^n\int^t_0dt_1...\int_0^{t_{n-1}}dt_n
\left(V(t_1)...V(t_n)\right)_c
$$
Here the index $c$  in $U_c$
indicates that one takes only the
connected diagrams. 

The similar relation is true for evolution operator with 
the adiabatic
cutoff.

Below  for simplicity of notations  we consider interactions with
the only one type of particles, but the main results are valid for 
arbitrary number of types of particles.

\section {Evolution operator
for non-translation invariant Hamiltonians}
\setcounter{equation}{0}
\subsection{Second order}
For the vacuum matrix element of the evolution operator we  obtain from 
(\ref{1a}) the
representation:
\begin{equation}
\langle0|U(t)|0\rangle=e^{{\cal E}(t)}
\label{L1}
\end{equation}
where
\begin{equation}
{\cal E}(t)=\langle 0|(-i\lambda\int^t_0dt_1V(t)+(-
i\lambda)^2\int^t_0dt_1\int^{t_1}_0dt_2V(t_1)V(t_2)+\dots)_c|0\rangle .
\label{L2}
\end{equation}
Here the symbol $(...)_c$ means that we keep only connected diagrams.

Representation (\ref{L1}), (\ref{L2}) 
permits us to calculate the leading terms of the asymptotic behaviour
of the matrix elements of the evolution operator
for large time $t$
as well the corrections to the leading terms.
In fact we will show that  ${\cal E}(t)$
has the following form
\begin{equation}
{\cal E}(t)=At+B+C(t)
\label{E2}
\end{equation}
where one has the perturbative expansions
\be
\label{EX1}
A=\lambda^2A_2+\lambda^3A_3+..., ~~~B=\lambda^2B_2+\lambda^3B_3+...,~~~
C(t)=\lambda^2C_2(t)+\lambda^3C_3(t)+...
\ee
and $C_n(t)$ vanishes for large $t$.

Let us find explicitly these terms in the 
second order of perturbation theory
for the Hamiltonian

\begin{equation}
H=H_0+\lambda V, \label{H1}
\end{equation}
where 
\be
\label{H2}
H_0=\int \omega (p)a^*(p)a(p)dp
\ee
and the interaction has the form
\be
V=\int (v(p_1,..., p_n)a^*(p_1)...a^*(p_n)+c.c.).
dp_1...dp_n
\label{3.30}
\ee
Here $\omega (p)$ is a positive smooth function, for example
$\omega(p)=\sqrt{p^2+m^2}, m>0$
and $v(p_1,...,p_n)$ is a   test  function.
For this interaction the first term in
(\ref{L2}) is identically zero.
The second term in (\ref{L2}) equals to

\be
\label{3.2}
{\cal E}^{(2)}(t)
= (-i\lambda)^2\int dp_1...dp_n |v(p_1,...,p_n)|^2
\int^t_0dt_1\int^{t_1}_0dt_2
e^{it_1E_1+it_2E_2}
\ee
where
\be
\label{3.3}
E_2=-E_1=E(p_1,...,p_n)=\sum_{i=1}^{n}\omega(p_i)
\ee

By using the equality
\be
\label{Exp1}
\int^t_0dt_1\int^{t_1}_0dt_2
e^{-it_1E+it_2E}=
-\frac{i}{E}t+\frac{1}{E^2}-\frac{1}{E^2}e^{-itE}
\ee
we get
\be
\label{EXP3}
{\cal E}^{(2)}(t)=
(-i\lambda)^2\int dp_1...dp_n |v(p_1,...,p_n)|^2
(-\frac{i}{E}t+\frac{1}{E^2}-\frac{1}{E^2}e^{-itE})
\ee

Therefore we obtain the expression of the form (\ref{E2})
\be
\label{EXP4}
{\cal E}(t)=\lambda^2A_2t+\lambda^2B_2+\lambda^2C_2(t)+...
\ee
where
\be
\label{A1_2}
A_2=i\int 
\frac{|v(p_1,...,p_n)|^2}{E(p_1,...,p_n)}dp_1...dp_n,
\ee

\be
\label{T2_2}
B_2=-\int 
\frac{|v(p_1,...,p_n)|^2}{E(p_1,...,p_n)^2}dp_1...dp_n,
\ee

\be
\label{T3_2}
C_2(t)=\int
\frac{|v(p_1,...,p_n)|^2}{E(p_1,...,p_n)^2}e^{-itE(p_1,...,p_n)}dp_1...dp_n
\ee

We have obtained the following

{\bf Theorem 1}. The 
vacuum expectation value of the evolution operator for
the Hamiltonian (\ref{H1}) in the second order of perturbation
theory has the form
\be
\label{as2}
<U(t)>=e^{\lambda^2A_2t+\lambda^2B_2+\lambda^2C_2(t)}
\ee
where $A_2,B_2$ and $C_2(t)$ are given by (\ref{A1_2}),(\ref{T2_2})
and (\ref{T3_2}).

{\bf Remark}.  By using
the stationary phase method one can prove that the function
$C_2(t)$  vanishes as $t\to\infty$ (see below).

\subsection{Decay}

We have proved theorem 1 under the assumption $\omega(p)>0$.
However the obtained formula (\ref{as2}) is valid in the
more general case when one has the decay.
In this case formula (\ref{as2}) still is true but
in the expressions (\ref{A1_2})-(\ref{T3_2}) one
has to substitute $E\to E-i0$. Let us consider the important case
when 
\be
\omega(p)={p^2\over 2}-\omega_0,~~\omega_0>0
\label{omega}
\ee
Instead of (\ref{Exp1}) we will use now the identity

\be
\label{Exp12}
\int^t_0dt_1\int^{t_1}_0dt_2
e^{i(t_2-t_1)E}=
t\int_0^t(1-{\sigma\over t})e^{-i\sigma E}d\sigma .
\ee
We have 
\be
\label{3.2S}
{\cal E}^{(2)}(t)=
(-i\lambda)^2\int dp_1...dp_n |v(p_1,...,p_n)|^2
\int^t_0dt_1\int^{t_1}_0dt_2
e^{i(t_2-t_1)E(p_1,...,p_n)}
\ee
$$
=-\lambda^2 t\int_0^td\sigma(1-{\sigma\over t})\int dp_1...dp_n 
|v(p_1,...,p_n)|^2e^{-i\sigma E(p_1,...,p_n)}
$$
$$
=\lambda^2 tA_2(t)+\lambda^2B_2(t)
$$
where
\be
\label{3.2T}
A_2(t)=
-\int_0^td\sigma F(\sigma),~~B_2(t)=\int_0^td\sigma\sigma F(\sigma)
\ee
and 
\be
\label{3.3TR}
F(\sigma)=
\int dp_1...dp_n 
|v(p_1,...,p_n)|^2e^{-i\sigma E(p_1,...,p_n)}.
\ee
By using the stationary phase method we obtain the following
asymptotic behaviour of the function $F(\sigma)$ as 
$\sigma\to\infty$:
$$
F(\sigma)=({2\pi i\over\sigma})^{dn/2}e^{in\sigma\omega_0}
|v(0)|^2[1+o({1\over\sigma})].
$$
Therefore for $dn\geq 3$ there exist the limits
\be
\label{3.3TP}
\lim_{t\to\infty}A_2(t)=A_2=-\int_0^{\infty}d\sigma F(\sigma)
\ee
$$
\lim_{t\to\infty}B_2(t)=B_2=\int_0^{\infty}d\sigma
\sigma F(\sigma)
$$
because there exists the limit
$$
\lim_{t\to\infty}\int_1^{t}e^{i\sigma\omega_0}{d\sigma\over
\sigma^{1/2}}.
$$
Moreover one has
$$
A_2(t)=-\int_0^{\infty}d\sigma F(\sigma)+o({1\over t^2}),~~
B_2(t)=\int_0^{\infty}d\sigma
\sigma F(\sigma)+o({1\over t}).$$
If $dn\geq 5$ one gets also 
\be
\label{3.3TI}
A_2=i\int dp_1...dp_n 
{|v(p_1,...,p_n)|^2\over E(p_1,...,p_n)-i0}
\ee

\be
\label{3.3TG}
B_2=-\int dp_1...dp_n 
{|v(p_1,...,p_n)|^2\over (E(p_1,...,p_n)-i0)^2}.
\ee
Indeed one has
$$
A_2=-\int_0^{\infty}d\sigma F(\sigma)=
-\lim_{\epsilon\to 0}\int_0^{\infty}d\sigma
F(\sigma)e^{-\sigma \epsilon}.
$$
This is true due to the Lebesgue theorem since 
$|F(\sigma)e^{-\sigma \epsilon}|\leq |F(\sigma)|$
and $F(\sigma)\in L_1(R_+)$ ($L_1$ is the  space of absolute integrable 
functions) if $nd\geq 3$. Substituting in the above formula the
representation (\ref{3.3TR}) and changing the order of integrations
(we can do this 
due to the Fubini theorem since for positive $\epsilon$
the function $|v(p_1,...,p_n)|^2e^{-i\sigma (E(p_1,...,p_n)-i\epsilon)}$
 belongs to the space $\L_1(R_+\times R^{nd})$ of 
absolute integrable functions), we can 
perform the integration over $\sigma$ explicitly
$$A_2=\lim_{\epsilon\to 0}(i) 
\int dp_1...dp_n 
{|v(p_1,...,p_n)|^2\over E(p_1,...,p_n)-i\epsilon}
=i\int dp_1...dp_n 
{|v(p_1,...,p_n)|^2\over E(p_1,...,p_n)-i0}.
$$
 The same calculation is true for $B_2$ with the more strong
 assumption : $dn\geq 5$,
$$
B_2=\int_0^{\infty}d\sigma\sigma F(\sigma)
=\lim_{\epsilon\to 0}\int_0^{\infty}d\sigma\sigma 
\int dp_1...dp_n 
|v(p_1,...,p_n)|^2e^{-i\sigma (E(p_1,...,p_n)-i\epsilon)}
$$
$$
=-\lim_{\epsilon\to 0} 
\int dp_1...dp_n 
{|v(p_1,...,p_n)|^2\over (E(p_1,...,p_n)-i\epsilon)^2}
=-\int dp_1...dp_n 
{|v(p_1,...,p_n)|^2\over (E(p_1,...,p_n)-i0)^2}.
$$
We have proved the following theorem.

{\bf Theorem 2}. The asymptotic behaviour as $t\to\infty$ of the 
vacuum expectation value of the evolution operator for
the Hamiltonian (\ref{H1})  with the dispersion low
(\ref{omega})
in the second order of perturbation
theory is 
\be
\label{as32}
<U(t)>=e^{\lambda^2A_2t+\lambda^2B_2+\lambda^2 o(1/t)}
\ee
where $A_2$ and $B_2$ are given by (\ref{3.3TP})
(or (\ref{3.3TI}) and (\ref{3.3TG})).
After the rescaling $t\to t/\lambda^2$ one gets the $\lambda^2$
corrections to the stochastic limit
\be
\label{as33}
<U(t/\lambda^2)>=e^{A_2t+\lambda^2B_2+\lambda^2 o(\lambda^2/t)}.
\ee

\subsection{Example}

We discuss here the evolution operator for the
simple explicitly solvable model described by  the Hamiltonian
\begin{equation}
H=\int\omega(k)a^+(k)a(k)d^d k+\lambda\int(a(k)\overline v(k)+
a^*(k)v(k))d^dk.
\label{(1)}
\end{equation}
We will see that the vacuum expectation value of the
evolution operator has the form obtained in theorems 1 and 2.
Under assumptions
\begin{equation}
\label{(2)}
 A=\lambda^2A_2=i\lambda^2\int
{|v(k)|^2\over\omega(k)}\,d^dk<\infty,~\quad
B=\lambda^2B_2=-\lambda^2\int{|v(k)|^2\over\omega^2(k)}\,d^dk<\infty\ ,
\end{equation}
one has the following
\bigskip

\noindent{\bf Proposition 1}. The vacuum expectation value of the
evolution operator $U(t)=e^{itH_0}$ $e^{-itH}$ is
\begin{equation}
\langle U(t)\rangle=\exp\left[At+
B+\lambda^2\int dk{|v(k)|^2\over\omega^2(k)}\,
e^{-i\omega(k)t}\right]
\label{(3)}
\end{equation}
\bigskip

\noindent{\it Proof\/}. It follows from the explicit solution
\be
\label{expl}
H=W^*(H_0+E_0)W,~~E_0=-\int{|v(k)|^2\over\omega(k)}dk.
\ee
$$
W=\exp\lambda\int{(a^*(k) v(k)-
a(k)\overline v(k))\over \omega(k)}dk=\exp\lambda\int
{a^*(k) v(k)\over \omega(k)}dk\exp-\lambda\int{
a(k)\overline v(k))\over \omega(k)}dk e^{ B/2}
$$
From Proposition 1 we obtain
\bigskip

\noindent{\bf Proposition 2}. The asymptotic behaviour of the
expectation value (\ref{(3)}) for $t\to\infty$ has the form
\begin{equation}
\label{(4)}
\langle U(t)\rangle=\exp\left[At+
B+\lambda^2\left({1\over t}\right)^{d\over2}\,(2i\pi)^{d
\over2}\frac{|v(k_0)|^2}{\omega^2(k_0)}\,e^{-i\omega(k_0)t}+...\right].
\end{equation}
where $k_0$ is a critical point, $\nabla\omega(k_0)=0$ and we assume
there is only one nondegenerate critical point.
\bigskip

\noindent{\bf Proof\/}. It follows immediately from (\ref{(3)}) 
by using the
stationary phase method.\bigskip

\noindent{\bf Remark 1\/}. After the rescaling $t\to t/\lambda^2$ we get
from (\ref{(4)}) corrections to the stochastic limit
\begin{equation}
\langle U(t/\lambda^2)\rangle=
\exp\left[ A_2t+
B+
\lambda^2\left({\lambda^2\over t}\right)^{d\over2}
(2i\pi)^{d\over2}{|v(k_0)|^2\over\omega^2(k_0)}\cdot
e^{-i\omega(k_0)t/
\lambda^2}+\dots\right].
\label{(5)}
\end{equation}

\bigskip

\noindent{\bf Remark 2\/}. One can take for example
$$v(k)=\omega(k)f(k)\ ,$$
$$\omega(k)={k^2\over2}\,-\omega_0\ ,\quad\omega_0>0$$
where $f(k)$ is a test function. 
Then the assumptions (\ref{(2)}) are satisfied,
$$B=-\lambda^2\int|f(k)|^2d^dk\ ,\quad A=i\lambda^2
\int\left({k^2\over2}\,-\omega_0
\right)|f(k)|^2d^dk\ ,$$
the critical point $k_0=0$ and (\ref{(5)}) has the form

$$\langle U(t)\rangle=\exp\left[At+
B+\lambda^2\left({1\over t}\right)^{d\over2}(2i
\pi)^{d\over2}|f(0)|^2e^{i\omega_0t}+\dots\right]$$

\bigskip

\noindent{\bf Remark 3\/}. If $\omega(k)=k^2-\omega_0$
and $v(k)$ is an arbitrary test function then one has the decay.
We can not use in this case the diagonalization of the Hamiltonian
(\ref{(1)}) but the formula (\ref{L1}) still is true. We have
$$
<U(t)>=\exp[- \lambda^2\int_0^tdt_1\int_0^{t_1}dt_2\int d^dk 
|v(k)|^2e^{i(t_2-t_1)\omega(k)}]
=\exp[\lambda^2tA_2(t)+\lambda^2B_2(t)]
$$
where
$$
A_2(t)=-\int_0^td\sigma F(\sigma),
~~B_2(t)=\int_0^td\sigma\sigma F(\sigma)
$$
and
$$
F(\sigma)=\int d^dk
|v(k)|^2e^{-i\sigma\omega(k)}
$$
By using this representation and the expansion
$$
F(\sigma)=({2i\pi\over \sigma})^{d\over2}|v(0)|^2
e^{i\sigma\omega(0)}(1+o({1\over\sigma}))
$$
we compute corrections to the stochastic limit. Notice also that the
stochastic limit
$$
\lim_{\lambda\to 0}\lambda^2\int_0^tdt_1\int_0^{t_1}dt_2
F(t_1-t_2)=\lim_{\lambda\to 0}\int_0^td\tau\int^0_{-\tau/\lambda^2}
d\sigma F(-\sigma)=t\int_0^{\infty}d\sigma F(\sigma)
$$
exists in any dimension $d=1,2,...$.

\subsection{Higher orders}

To formulate the theorem about the asymptotic behaviour of 
the vacuum expectation value
of the evolution operator
it is convenient to use the  Friedrichs
$\Gamma $ - operation

\be
\label{H7}
\Gamma (V)=\lim _{\epsilon \to +0}(-i)\int _0^{\infty}
e^{-\epsilon t}e^{itH_0}Ve^{-itH_0}dt
\ee
acting to the monomials as

\be
\Gamma (V_{I,J})=
\int (\Gamma v)(p_1,i_1\dots p_{I},i_I|q_1,j_1\dots q_J,j_J)
\prod^I_{l=1}a^*_{i_l}(p_l) dp_l\prod^J_{r=1}a_{j_r}(q_r)dq_r
\label{H4}
\ee
where

$$
(\Gamma v)(p_1,i_1\dots p_{I},i_I|q_1,j_1\dots q_J,j_J)=
\frac{v(p_1,i_1\dots p_{I},i_I|q_1,j_1\dots q_J,j_J)}
{\sum _{l=1}^I \omega _{l}(p_l)-\sum _{r=1}^J\omega _{r}(p_r)+
i0}.$$
It has  the property

\be
\label{H6}
[H_0,\Gamma _{\pm}(V)]=V.
\ee

The following theorem  gives a very  effective exact representation
 for  the vacuum expectation value
of the evolution operator.

{\bf Theorem 3.} Vacuum expectation value of the evolution
operator for the Hamiltonian (\ref{4}) has the following representation
\begin{equation}
\label{L3}
<U(t)>=e^{At+B+C(t)}
\end{equation}
where  

\begin{equation}
A=i\sum _{n=2}^{\infty} 
\lambda ^n\langle0| (\underbrace{V\Gamma(V...\Gamma(V)...))}_{n})_c
|0\rangle,~~~
\label{L4a}
\ee
\be
\label{L4b}
B=\sum_{n=2}^{\infty} \lambda ^n\sum_{l=2}^{\infty}(-1)^{l-1}
\sum_{k_1+...+k_l=n}
\langle0| (W_{k_1}...W_{k_l})_c|0\rangle,
\end{equation}
and
\be
\label{Gam}
W_n=\underbrace{\Gamma(V...\Gamma(V)...)}
_{n}
\ee
The function $C(t)$ is given 
by the sum of terms of the form
\be
\label{CT}
\int dp...e^{it\sum _{l=1}^{k_1}E_k}
\underbrace{\Gamma (v_{i1}\circ ...\Gamma(v_{i2}\circ )...)}_{k_1}
\underbrace{\Gamma (v_{i3}\circ ...\Gamma(v_{i4}\circ )...)}_{k_2}...
\underbrace{\Gamma (v_{i5}\circ ...\Gamma(v_{i6} )...)}_{k_{l+1}}(p,...).
\ee

To prove this Theorem let us consider one of the terms representing 
the  n-th order  of the
perturbation theory with adiabatic parameter $\epsilon >0$

\be
\label{3.7}
\int dp ...{\cal E}^{(n)}_0 (p,...)=
\ee
$$(-i\lambda)^n\int dp ...
\int^{t}_0dt_1\int^{t_1}_0 dt_2...\int^{n-1}_0dt_n
e^{\sum _k (iE_kt_k-\epsilon t_k)}(v_{i_1}\circ 
...\circ v_{i_n})(p...),
$$

\begin{figure}[h]
\begin{center}
\epsfig{file=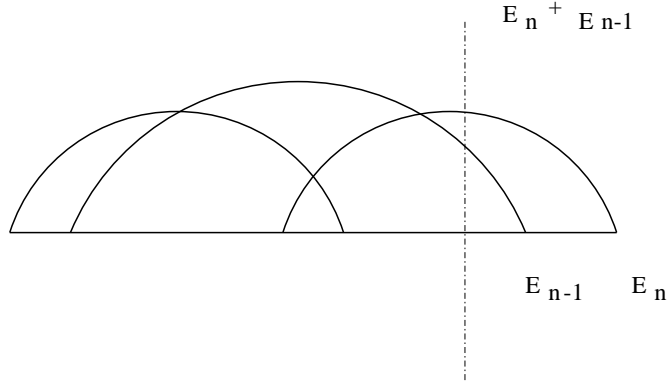,
   width=250pt,
   angle=0
 }
\end{center}
\caption{Diagram with energy-cuts}
\label{Fig-cut}
\end{figure}

Integrating over $t_n$ one gets two terms. One contains $e^{iE_nt_{n-1}}$
and the other contains $1$. Together with the factor coming from 
$V(t_{n-1})$ the exponent 
 $e^{iE_nt_{n-1}}$ yields to $e^{i{\cal E}_{n-1}t_{n-1}}$, where
${\cal E}_{n-1}=E_n+E_{n-1}$ is the total energy of the lines 
crossed by $n-1$-cut (see Fig.\ref{Fig-cut}).

 Therefore, the contribution coming from the upper bounds of
integration over 
$t_n, ...   t_2$
is equal to $e^{i{\cal E}_{2}t_{1}}$, where
${\cal E}_{2}=\sum _{k=2}^nE_k.$
It is evident that ${\cal E}_{2}=-E_1$ and therefore the integrand does not depend on
$t_1$ and intergration of
this terms over $t_1$ produces 
\be
t\int dp ... (\underbrace{v\circ  
\Gamma(v\circ  ...\Gamma(v\circ  )...)}_{n} (p....)
\label{kasy}
\ee
All these terms correspond to 
\be
t\langle0| (\underbrace{V\Gamma(V...\Gamma(V)...)}_{n})_c|0\rangle .
\label{asy}
\ee
Or, in other words, we  have proved that contributions from the 
upper bounds produce the linear in $t$ terms. 
To prove that the term (\ref{asy}) gives the leading term 
in asymptotic expansion
we have to prove that the rest terms are of order 1 when 
$t\to \infty$.
Or in other words we have to prove that the terms which contain at least 
one 
contribution 
from the down bound are of order 1.

To make more clear the proof we present in Appendix A an explicit
calculation of ${\cal E}_0$ up to the fourth order 
of the perturbation theory.

Let us first consider contribution 
from only one the lower bound. 
For example, the integration over $t_1$ of  the term produced by  the
low bound of integration over $t_{k+1}$ and all other upper bounds
gives:
$$
\int dp ...\int _0^{t_1}dt_1e^{i\sum _{l=1}^kE_l t_1}
 (\underbrace{v\circ  \Gamma(v\circ  ...\Gamma(v\circ  )...)}_{k}
\underbrace{\Gamma(v\circ  ...\Gamma(v )...)}_{n-k})(p...) =
$$
\be
i\int dp ...[e^{it\sum _{l=1}^kE_l}-1](\underbrace
{\Gamma (v\circ  \Gamma(v\circ  ...\Gamma(v\circ  )...))}_{k}
\underbrace{\Gamma(v\circ  ...\Gamma(v  )...)}_{n-k})(p...).
\label{asyy}
\ee
We see that (\ref{asyy}) gives a contribution to terms in (\ref{L2})
that do not depend on $t$. All these contributions  from
the down bound of integration over $t_k$ and $t_1$ and all other upper 
bounds sum up to the following expression

\be
-\langle0| (\underbrace
{\Gamma V (\Gamma(V ...\Gamma(V )...))}_{k}
\underbrace{\Gamma(V ...\Gamma(V )...)}_{n-k})_c|0\rangle .
\ee

Let us note the obvious difference between 
\be
\label{GGam}
\underbrace{\Gamma(V...\Gamma(V)...)}_{k}
\underbrace{\Gamma(V...\Gamma(V)...)}_{l}
~~\mbox{and }~~~
\underbrace{\Gamma(V...\Gamma(V)...)}_{k+l}.
\ee

\begin{figure}[h]
\begin{center}
\epsfig{file=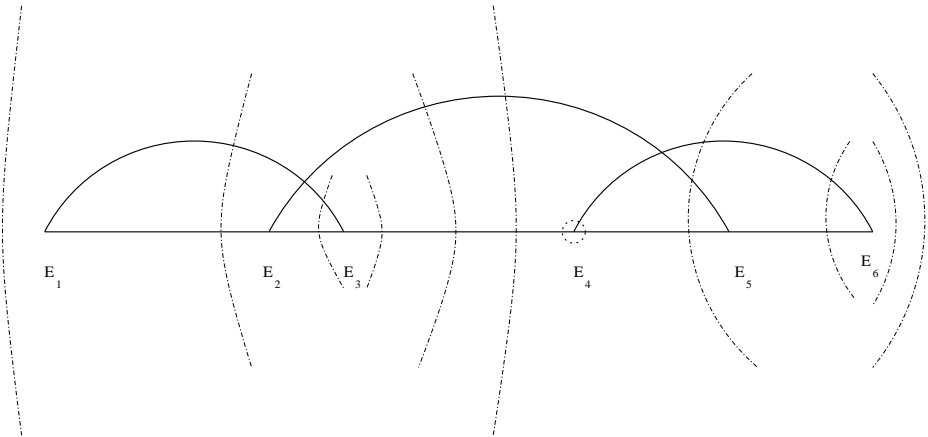,
   width=350pt,
  angle=0}
\end{center}
\caption{Diagram corresponding to $\Gamma (V\Gamma(V\Gamma(V)))
\Gamma (V\Gamma (V\Gamma(V)))$}
\label{Fig1}
\end{figure}

This expression can be
presented on graphs, see Fig.\ref{Fig1}. On the graph
the directions of applications 
of $\Gamma $ operators are indicated by 
brackets. The term in (\ref{asyy}) with $e^{it\sum _{l=1}^kE_l}$
produces the decreasing contribution as $t \to \infty$ since 
we have $\sum _{l=1}^kE_l>0$.

For the n-th order of perturbation theory
the terms corresponding to contributions of $l$ lower bounds
have the following structure 
\be
\label{n-or}
\int dp...[e^{it\sum _{k=1}^{k_1}E_k}-1]
\underbrace{\Gamma (v\circ ...\Gamma(v\circ )...)}_{k_1}
\underbrace{\Gamma (v\circ ...\Gamma(v\circ )...)}_{k_2}...
\underbrace{\Gamma (v\circ ...\Gamma(v )...)}_{k_{l+1}}(p,...)
\ee
$\sum _{i=1}^{l+1}k_i=n.$ All these terms either contain oscillator factors
or give the constant factors. 
The constant factors can be sum up to  expressions

\be
\label{cf}
(-)^l <0|(
\underbrace{\Gamma (V ...\Gamma(V )...)}_{k_1}
\underbrace{\Gamma (V ...\Gamma(V )...)}_{k_2}...
\underbrace{\Gamma (V ...\Gamma(V )...)}_{k_{l+1}})_c|0>.
\ee

Therefore we have shown that 
\begin{equation}
{\cal E}_0=tA +B+C(t),\label{L3B}
\end{equation}
where $A$ is given in perturbation theory by (\ref{L4a}) 
and $B$ is given by
\begin{equation}
B=\sum _n \lambda ^n \sum_l\sum_{k_1+...k_l=n}
(-1)^{l-1}\langle0|(\underbrace{\Gamma (V ...\Gamma(V )...)}_{k_1}
\underbrace{\Gamma (V ...\Gamma(V )...)}_{k_2}...
\underbrace{\Gamma (V ...\Gamma(V )...)}_{k_{l}})_c|0\rangle .
\label{L4aa}
\end{equation}

The function $C(t)$ is the sum of terms of the form

\be
\label{fff}
\int dp...e^{it\sum _{l=1}^{k_1}E_k}
\underbrace{\Gamma (v\circ ...\Gamma(v\circ )...)}_{k_1}
\underbrace{\Gamma (v\circ ...\Gamma(v\circ )...)}_{k_2}...
\underbrace{\Gamma (v\circ ...\Gamma(v )...)}_{k_{l+1}}(p,...).
\ee
These terms vanish as $t \to \infty$ because $\sum _{l=1}^{k_1}E_k
\neq 0$.

\subsection{Wave operators and the main formula}

In this section we show how the spectral theory and renormalized 
wave operators  can be used for the derivation 
of the main formula.
In particular explicit expressions for the parameters $A,B$ and $C(t)$
will be obtained.

We consider the Hamiltonian (\ref{4}) for one type of particles
with $\omega(p)=\sqrt{p^2+m^2}, m>0$
in the space $R^d, d>2$. We will work with the formal perturbation
series for the evolution operator. In fact if the interaction
(\ref{4}) includes only fermionic operators or it is linear
in bosonic operators then one can prove that the series are absolutely
convergent. The following operator plays the crucial role in the 
scattering theory 
\be
\label{Tfor}
T=:\exp\sum_{n=1}^{\infty}(-\lambda)^n
\Gamma(V...\Gamma(V)...)_L:
\ee
Here $()_L$ means that only connected non-vacuum diagrams are included.
The operator $T$ is equal in fact to the non-vacuum part
of the conjugate wave operator :
$$
T=\lim_{\epsilon\to 0}\lim_{t\to\infty}U^*_{\epsilon}(t)/<U^*_{\epsilon}(t)>
$$
One has the following relations 
\cite{Hepp}:
\be
\label{Diag}
HT=T(H_0+E_0),
\ee
$$
E_0=\sum_{n=1}^{\infty}(-\lambda)^{n+1}<V\Gamma(V...\Gamma(V)...)_c>,
$$
$$T^*T=TT^*=Z^{-1},
$$
$$
Z^{-1}=||T\Phi_0||^2
$$
We will use these relations to derive the main formula for matrix elements
of the evolution operator and in particular to compute corrections
to the stochastic limit.

From (\ref{Diag}) it follows
$$
H=T(H_0+E_0)T^*e^B
$$
where
\be
\label{Bfor}
B=\ln Z
\ee
Therefore one has
\be
\label{U(t)}
U(t)=e^{itH_0}e^{-itH}=e^{itH_0}Te^{-it(H_0+E_0)}T^*e^B
=e^{At+B}e^{itH_0}Te^{-itH_0}T^*
\ee
where
\be
\label{Afor}
A=-iE_0
\ee
By taking the expectation value of the equality (\ref{U(t)})
we obtain
\be
\label{psi1}
<\psi,U(t)\psi>=e^{At+B+C(t)}
\ee
where
$$
e^{C(t)}=<\psi,e^{itH_0}Te^{-itH_0}T^*\psi>
$$
If $\psi$ is the vacuum vector then one can prove that $C(t)\to 0$ as 
$t\to\infty$ and we obtain the main formula (\ref{L3}). If $\psi$
is a non-vacuum vector then the asymptotic behaviour of $C(t)$
is more complicated. We have proved the following theorem.

{\bf Theorem 4}. If the Hamiltonian satisfies 
the indicated above assumptions 
then there exists the following 
representation for the vacuum expectation
value of the evolution operator
$$
<\Phi_0,U(t)\Phi_0>=e^{At+B+C(t)}
$$
where constants $A$ and $B$ are given by (\ref{L4aa}) and (\ref{L4b})
and $C(t)$ is defined as
\be
e^{C(t)}=<\Phi_0,T(t)T^*\Phi_0>,~~
T(t)=e^{itH_0}Te^{-itH_0}
\label{LC}
\ee
here the weak limit of $T(t)$ as $t \to \infty$ 
is equal to 1 and $\lim_{t\to\infty}C(t)=0$.

This theorem is closely related with the theorem 3 and it shows
the physical meaning of constants $A,B$ and the function $C(t)$.

\section {One particle  matrix element of the evolution operator for
translation invariant Hamiltonian}
\setcounter{equation}{0}
In this section we study the asymptotic behaviour of one 
particle matrix elements of evolution operator 

\be
\label{U1}
<p|U(t,\lambda)|p'>= \delta(p-p')U_{1,1}(t,p,\lambda),
\ee
for translation invariant Hamiltonian (\ref{H8}) without 
vacuum polarization, i.e. when $V_{I0}=V_{0J}=0$, and we assume also 
$V_{1,1}=0$.
We will prove the following 

\bigskip

{\bf Theorem 5.}
For $t\lambda^2$=fixed =$\tau $ and small $\lambda^2$
one has the following representation

\be
U_{1,1}(t,p,\lambda)=
\exp \{i\tau A_2(p)\}
(1+i\lambda ^2\tau A_4(p)+\lambda ^2B_2(p) +o(\lambda^2))\label{N7}
\ee
where
\be
A_2(p)=(V\Gamma(V))_{1,1}(p)
\label{N11}
\ee
\be
B_2(p)=
(V\Gamma^2(V))_{1,1}(p)
\label{N13}
\ee
$$
A_4(p)=A_4^{1PI}(p)+A_4^{1PR}(p),
$$

\be
A_4^{1PI}(p)=(V\Gamma(V\Gamma(V\Gamma(V))))^{1PI}_{1,1}(p)
\label{N12}
\ee

\be
A_4^{1PR}(p)=(V\Gamma(V)(V\Gamma^2(V))))^{1PR}_{1,1}(p)
\label{N12R}
\ee

Here  subscript $_{1,1}$ means that we left only connected diagrams
in the corresponding expression with one creations and one 
annihilation operator 
and  we take only the 
coefficient in front of $\delta(p-p')$,
\be
<p|{\cal O}|p'>={\cal O}_{1,1}(p)\delta(p-p')
\ee

\bigskip

{\bf Remark 1}.
We will give the proof of Theorem 5  for the Hamiltonian of the form
\be
\label{ham}
H=\int \omega (p)a^+(p)a(p)dp+
\lambda \int (v(p,q)a^+(q)a^+(p-q)a(p)+c.c.)dpdq.
\ee
Generalization to the Hamiltonians in the general form  (\ref{H8})
is obvious and the suitable comments will be given below.

\bigskip

{\bf Remark 2}. In fact the following representation is true
\be
U_{1,1}(t,p,\lambda)=
\exp \{ itA(p,\lambda)+B(p,\lambda)\}\left(1+C(t,p,\lambda)\right)
\label{N7a}
\ee
where $A(p,\lambda)$, $B(p,\lambda)$ and $C(t,p,\lambda)$
are formal series in $\lambda$ 
\be
A(p,\lambda) =\sum _{n=1}^{\infty}\lambda^{n}A_n(p),~~
B(p,\lambda) =\sum _{n=1}^{\infty}\lambda^{n}B_n(p),~~
C(t,p,\lambda) =\sum _{n=1}^{\infty}\lambda^{n}C_n(t,p)
\ee
and functions  $C_n(t,p)$ vanish as $t \to \infty$.
In the case of $<p|V|p'>=0$ ,
$A_1=0$, , $B_1=0$ and 
low order terms in these series are given by
(\ref{N11}), (\ref{N13}) and 
\be
C_2(t,p)=(e^{itH_0}Ve^{-itH_0}\Gamma^2(V))_{1,1}(p).
\label{N14}
\ee
We will give the proof of (\ref{N7}) in the next section 
using the methods of scattering theory. This proof assumes that  
one has  not decay. 

\bigskip

{\bf Remark 3.} Representation (\ref{N7}) means a special dependence
of  $U_{1,1}(t,\lambda,p)$ on $t$. Namely it says that (2n)-order
on $\lambda$ of $U_{1,1}(t,\lambda,p)$
$$U_{1,1}(t,\lambda,p)=\sum _n
\lambda ^{2n}U^{(2n)}_{1,1}(t,p),$$
can be represent  as 
\be
U^{(2n)}_{1,1}(t,p)=\sum _{k=0}^n t^k~U^{(2n,k)}_{1,1}(p)
~+~
r_n(t,p)
\ee
with $r_n(t)\to 0$ for $t\to \infty$.

\bigskip

{\bf Remark 4.} Performing the
stochastic limit we get 
\be
\label{st1}
U_{1,1}(\tau/\lambda ^2,p)\to \exp i\tau A_2(p)~~\mbox{as }~~\lambda \to 0,
\ee
i.e. the stochastic limit of one particle matrix elements of evolution
 operator, 
$U(\tau/\lambda ^2)$, 
for translation invariant Hamiltonian without vacuum polarization
is given by the simple second order diagram (Fig.\ref{Fig4-or}).
This result takes place for arbitrary interaction (\ref{H8}) and the 
answer can be representing in the form

\be
\label{st2}
U_{1,1}(\tau /\lambda ^2,p)\to 
\exp\{i\tau (V\Gamma(V))^{(c)}_{1,1}(p)\}~~\mbox{as }~~\lambda \to 0.
\ee
Here the symbol $(...)^{(c)}_{1,1}$ means that we left only connected 
diagrams
in $V\Gamma(V)$ with one creations and one annihilation operator 
corresponding to the same type of particles.

\begin{figure}[h]
\begin{center}
\input{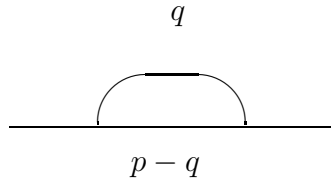}
\end{center}
\caption{The second order mass renormalization diagram } 
\label{Fig2-or}\end{figure}

\bigskip

\begin{figure}[h]
\begin{center}\epsfig{file=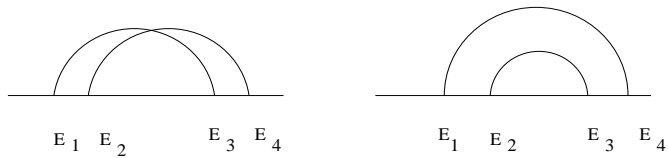,
   width=250pt,
   angle=0}
\end{center}
\caption{The 4-order 1PI mass renormalization diagrams } 
\label{Fig4-orI}
\end{figure}

{\bf Remark 5.} There are two sources of corrections to the 
stochastic limit.
One  is the terms  in the second order approximation 
of $U^{(2)}_{1,1}(t)$ which does not depend on $t$,
 i.e. it corresponds to diagram 
Fig.\ref{Fig4-or} .
There are also corrections to the leading terms of the stochastic limit
which come from  4-th order terms with linear dependence
on $t$. These corrections come from the 4-th order 1PI 
diagrams such as shown on 
Fig.\ref{Fig4-orI}.  (see for details Appendix  where we show  that  
contributions of these diagrams include the terms proportional to $t$ and 
after rescaling
produce a factor $\tau \lambda^2$).

\be
\label{st11}
U_{1,1}(\tau /\lambda ^2,p)\to \exp \{i\tau A_2(p)\}
[1+\lambda ^2(i\tau A_4(p)it+B_2(p))]
~~\mbox{as }~~\lambda \to 0.
\ee

\bigskip

{\bf Remark 6.} Let us note that $A_2(p)$ in (\ref{N11})
is positive  in the case when we assume that there is no decay
in the model. For Hamiltonian (\ref{ham}) the explicit expression for 
$A_2(p)$ is

\be
\label{exA2}
A_2(p)=\int dq (\overline v\circ\Gamma v)(p,q),
\ee
where
\be
\label{cirl}
(\overline v\circ\Gamma v)(p,q)=\frac{|v(p,q)|^2}
{\omega(p-q)+\omega(q)-\omega(p)}
\ee
We see that for the typical example $\omega(q)=\sqrt{q^2+m^2}$
the denominator in (\ref{cirl}) does not vanish.
In the case of different particles, say, 3 different particles
(Lee model)

\be
\label{ham_3p}
H=\int \omega_a (p)a^+(p)a(p)dp+
\int \omega_b (p)b^+(p)b(p)dp+\int \omega_c (p)c^+(p)c(p)dp~+
\ee
$$
\lambda \int dpdq( v(p,q)(a^+(q)b^+(p-q)c(p)+c.c.)
$$
we have the $\epsilon$-prescription for the denominator
\be
\label{cirla}
A_2(p)=\int dq~\frac{|v(p,q)|}
{\omega_a(p-q)+\omega_b(q)-\omega_c(p)+i\epsilon}
\ee
and if one has  a decay then  $A_2(p)$ gets the imaginary part.

\bigskip

{\bf Proof.} Now let us prove the Theorem 5. 
Let us  first make calculations up to the second order
\footnote{ An explicit calculation
of $U_{1,1}(t,p,\lambda)$
up to 4-th order is presented in Appendix B
}.
There is one  the
second-order diagram (Fig.\ref{Fig2-or}) and it gives :
 
$${\cal U}_2(t,p,q)=(-i\lambda)^2\int^t_0dt_1\int^{t_1}_0dt_2|v(p,q)|^2
e^{iE_2(-t_1+t_2)}=$$

\be
\label{U_2}
it\lambda^2(\overline v\circ\Gamma v)(p,q)- \lambda^2(\overline v
\circ\Gamma^2 v)(p,q)(1-e^{iE_1t}) 
\ee
where $E_1=-E_2, E_2=\omega(k)+\omega(p-k)-\omega(p)$. Here  and below 
for
simplicity of writing only the kernel
${\cal U}_{1,1}(T,p,q_1,...)$ of $U_{1,1}(t,p)$ is written,

\be
U^{(2n)} _{1,1}(t,p)=\int dq_1...dq_r{\cal U}_{2n}(t,p,q_1,...q_r).
\ee
In the case of a general Hamiltonian instead of (\ref{U_2}) one has 
\be
U^{(2)} _{1,1}(t,p)=it\lambda^2(V\Gamma (V))_{1,1}(p)
-\lambda^2(V\Gamma ^2(V))_{1,1}(p)+
\lambda^2(e^{iH_0t}Ve^{-iH_0t}\Gamma ^2(V))_{1,1}(p).
\ee

\begin{figure}
\begin{center}
\input{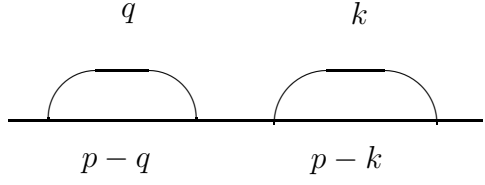}
\end{center}
\caption{Reducible mass renormalization diagram with two 
irreducible mass insertions}
\label{Fig5} 
\end{figure}

\begin{figure}[h]
\begin{center}\epsfig{file=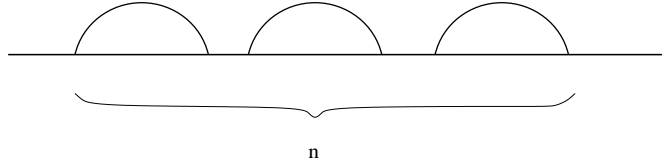,
   width=250pt,
   angle=0
 }
\end{center}
\caption{ 1PR mass renormalization diagram with n irreducible 2-order
insertions } 
\label{Figna}
\end{figure}

We will give the proof 
by induction. We already has proved the claim of the theorem at the second
order of perturbation theory.
Let us suppose that the 1PR diagrams including $n$ irreducible 
second order
mass insertions (see Fig.\ref{Figna})
have the following behaviour

\be
\label{beh_n1}
U^{1P(n)R}_{1,1 ~(2n)}(t)(p)=
\frac{(it)^{n}}{n!} A^n_2(p)+
\frac{n(it)^{n-1}}{(n-1)!} B_2(p)
A^{n-1}_2(p)
+....]
\ee
Here down index means the order in $\lambda$
and index $n$ in upper script  means the number
of irreducible insertions and dots means low order on T.

\begin{figure}[h]
\begin{center}\epsfig{file=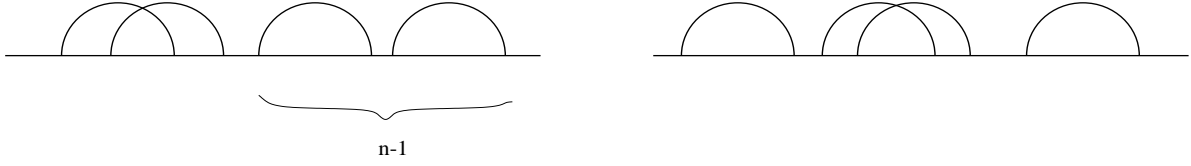,
   width=450pt,
   angle=0
 }
\end{center}
\caption{ 1PR mass renormalization diagram with (n-1) irreducible 2-order
insertions and one irreducible 4-order
insertion} 
\label{Fignb}
\end{figure}

We also assume that the 1PR diagrams  which include  $(n-2)$ irreducible 
second order
mass insertions  and one irreducible 4-th order mass insertion
(see Fig.\ref{Fignb}) have the following behaviour

\be
\label{beh_n}
 U^{1P(n-1)R}_{2n}(t)(p)=
\frac{(it)^{n-1}}{(n-2)!}
 A^{1PI}_4(p) A^{n-2}_2(p)+....
\ee

1RR diagrams 
with (n+1) irreducible second order
mass insertions (see Fig.\ref{Figna})
have the following representation

\be
\label{rep_2}
U^{1P(n+1)R}_{2(n+1)}(t,p)=
(-i)^{2}\int^t_0dt_1
\int^{t_1}_0dt_2\int dq|v(p,q)|^2
e^{iE_2(-t_1+t_2)} U^{1P(n)R}_{2n}(t_2,p),
\ee
Using the assumption (\ref{beh_n1}) and 

\be
\label{asT}
\int^t_0dt_1
\int^{t_1}_0dt_2
e^{iE(-t_1+t_2)}t^n_2=\frac{t^{n+1}}{(n+1)(iE)}
-\frac{t^{n}}{(iE)^2}+\frac{n(n-1)}{n+1}\frac{t^{n-1}}{(iE)^3}+
...
\ee
we see that the RHS of (\ref{rep_2}) has the following behaviour

$$(-i)^{2}\int^t_0dt_1
\int^{t_1}_0dt_2\int dq |v(p,q)|^2
e^{iE_2(-t_1+t_2)}
~[~\frac{(it_1)^{n}}{n!}
 A^n_2(p)+
\frac{n(it_1)^{n-1}}{(n-1)!} B_2(p)A^{n-1}_2(p)+...]
=
$$

$$
(i)^{2}\int dq \{\frac{(it)^{n+1}}{i(n+1)!}\frac{|v(p,q)|^2}{iE_2}A^n_2(p)
-\frac{(it)^{n}}{n!}\frac{|v(p,q)|^2}{(iE_2)^2}A^n_2(p)
+n\frac{(it)^{n}}{in!}\frac{|v(p,q)|^2}{iE_2} B_2(p,q)A^{n-1}_2(p)
+....\},
$$
i.e. the contribution of 1RR diagrams 
with $(n+1)$ irreducible second order
parts (see Fig.\ref{Figna})
has the following asymptotic

\be
\label{U2n}
 U^{1P(n+1)R}_{2(n+1)}(t,p)=
\frac{(it)^{n+1}}{(n+1)!}A^{n+1}_2(p)+
(n+1)\frac{(it)^{n}}{n!} B_2(p)
A^n_2(p)+...
\ee
 Here 
\be\label{AB_2}
A_2(p)=\int dq\frac{|v(p,q)|^2}{E_2},~~~
B_2(p)=-\int dq\frac{|v(p,q)|^2}{E_2^2},
\ee
or in the more general case
\be\label{AB_2G}
A_2(p)=(V\Gamma(V))_{1,1}^{1PI}(p),~~~
B_2(p)=-(V\Gamma^2(V))_{1,1}^{1PI}(p)
\ee
Therefore, (\ref{U2n}) is  in accordance with (\ref{beh_n1}), i.e.
(\ref{beh_n1}) is proved by induction.

1PR diagrams included  $(n-1)$ irreducible second order
mass insertions   and one irreducible 4-th order mass insertion
 (see Fig.\ref{Fignb})
have the following representation
\be
\label{rep4}
U^{1P(n)R}_{2(n+1)}(t,p)=
(-i)^{2}\int^t_0dt_1
\int^{t_1}_0dt_2\int dq |v(p,q)|^2
e^{iE_2(-t_1+t_2)}U^{1P(n-1)R}_{2n}(t_2,p)+
\ee
$$
(-i)^{4}\int^t_0dt_1
\int^{t_1}_0dt_2\int^{t_2}_0dt_3\int^{t_3}_0dt_4
\int dq dq_1
(\overline v\circ\overline v\circ v\circ v)_{1,1}^{(1PI)}(p,q,q_1)
~e^{i\sum _{i=1}^{4}E_it_i} ~U^{1P(n-1)R}_{2(n-1)}(t_4,p)
$$

Using
 \be
\label{asT4}
\int^t_0dt_1
\int^{t_1}_0dt_2\int^{t_2}_0dt_3\int^{t_3}_0dt_4~t^n_{4}
~e^{i\sum _{i=1}^{4}E_it_i} =\frac{1}{i^3}\frac{t^{n+1}}{n+1}\frac{1}{
(E_4)(E_4+E_3)(E_4+E_3+E_2)},
\ee
(\ref{asT}) and the assumptions (\ref{beh_n1}) and (\ref{beh_n})
we see that the RHS of (\ref{rep4}) has the following behaviour

$$
(i)^{2}\int^t_0dt_1
\int^{t_1}_0dt_2\int dq|v(p,q)|^2
e^{iE_2(-t_1+t_2)}
[\frac{(it_2)^{n-1}}{(n-2)!}
 A^{1PI}_4(p) A^{n-2}_2(p)+....]+
$$

$$
(i)^{4}\int^t_0dt_1
\int^{t_1}_0dt_2\int^{t_2}_0dt_3\int^{t_3}_0dt_4
dqdq_1(\overline v\circ\overline v\circ v\circ v)_{1,1}^{1PI}(p,q,q_1)
e^{i\sum _{i=1}^{4}E_it_i}~[~
\frac{(it_4)^{n-1}}{(n-1)!}A^{n-1}_2(p)+...]=
$$

$$
=\frac{(it)^{n}}{(n-2)!n}\int dq(\overline v\circ 
\Gamma(v))_{1,1}^{1PI}(p,q)A^{1PI}_4A_2^{n-2}
+i\frac{(it)^{n}}{n!}\int dqdq_1(\overline v\circ\Gamma(\overline
v\circ \Gamma(v\Gamma(v))))_{1,1}^{1PI}(p,q,q_1)A_2^{n-1},$$
Here 

\be
\label{A_4}
A^{1PI}_4(p)=\int dqdq_1(\overline v\circ\Gamma(\overline
v\circ \Gamma(v\Gamma(v))))_{1,1}^{1PI}(p,q,q_1)
\ee
or more generally,
\be
\label{A_4G}
A^{1PI}_4(p)=(V\Gamma(V \Gamma(V\Gamma(V))))_{1,1}^{1PI}(p)
\ee
Therefore,
$$
U^{1P(n-1)R}_{2(n+1)}(t,p)=\frac{(it)^{n}}{(n-1)!}tA_4(p)A_2^{n-1}(p),
$$
that is in accordance with assumption (\ref{beh_n}). 

It is evident that the behaviour (\ref{beh_n1}) and (\ref{beh_n}) implies 
(\ref{N7}) with
$A_2$  and $B_2$ as in the formulation of Theorem 5.

\bigskip

\subsection{Wave operators and the main formula}

In this section we show how the spectral theory and renormalized wave
operators  can be used for 
the derivation 
of the main formula.
In particular, explicit recursive  relations 
for the parameters $A_n,B_n$ and $C_n(t)$
will be obtained.
 Note that for 
this derivation we have to assume that the is no
 decay.

The intertwining operator $T$ is defined as a solution of the following
equation 
\be
\label{Inter}
HT=T(H_0+M),
\ee
Here $M$ has the form 
\be
\label{mex}
M=\int m(p)a^*(p)a(p)dp
\ee
The operator $T$ plays the crucial role in the scattering
theory. Its singular part defines the renormalized wave operators.
The renormalized wave operators
also give a solution of intertwining condition.
Taking $T$ in the following form 
\be
\label{TW}
T=:\exp W:, ~~~W=\Gamma(Q)
\ee
one gets \cite{ARE}   equations to define $Q$ and $M$.
\be
\label{T}
Q+V\zz T-(V\zz T)_{1,1}-W\cc M=0
\ee
\be
\label{M}
M=(V\zz T)_{1,1}
\ee
Here the symbol $\zz$ means that for connected $A$ in the 
operators $A\zz B$  all connected parts
of B are paired with $A$. If $B$ is connected then $A\zz B=
A\cc B=(AB)_c$.
For a special form of the interaction, when $V_{0I}=V_{I0}=0$
we can write  $M=(V\zz T)_{1,1}=(V\Gamma(Q))_{1,1}$.

Expanding $M$ and $Q$ in the power series in $\lambda$
\be
\label{se}
M=\sum _{n=1} \lambda ^{2n} M_{2n},~~
Q=\sum _{n=1} \lambda ^{2n+1} Q_{2n+1},
\ee
we get recursive relations to define $M_{2n}$ and $Q_{2n+1}$.
Let us compute explicitly the first terms solving these equations.
We obtain
\be
\label{Q1}Q_1=-V
\ee
\be
\label{M2}
M_2=-(V\Gamma V)_{1,1}
\ee
\be
\label{Q2}
Q_2=(V\Gamma V)_{c}-(V\Gamma V)_{1,1}
\ee
\be
\label{Q3}
Q_3=
-(V\Gamma _r(V \Gamma (V)))_c-\frac{1}{2}V\zz :\Gamma (V)^2:+
\Gamma V\cc M_{2}
\ee
\be
\label{M4}
M_4=
-(V\Gamma(V\Gamma _r(V \Gamma (V))))_{1,1}+
(V\Gamma ^2(V)M_{2})_{1,1}
\ee
or 
\be
\label{M42}
M_4=
-(V\Gamma(V\Gamma _r(V \Gamma (V))))_{1,1}
-(V\Gamma ^2(V)(V \Gamma (V))_{1,1})_{1,1}
\ee

Here we use the notation
\be
\label{rg}
\Gamma _r(Q)=\Gamma (Q-Q_{1,1})
\ee
One can  construct in perturbation theory the operator $Z$ such that
\be
\label{T*}
TT^*Z=1
\ee
In the second order in $\lambda$
\be
<p|Z|p'>=(1 + \lambda ^2B_2(p))\delta (p-p')
\ee
Therefore one has
\be
\label{U(t)11}
U(t)=e^{itH_0}e^{-itH}=e^{-itM}e^{it(H_0+M)}Te^{-it(H_0+M)}T^*Z
\ee
By taking the one particle expectation value of the 
equality (\ref{U(t)11})
\be
\label{ABc}
<p|U(t)|p'>=e^{-iM(p,\lambda )t}Z(p,\lambda )(1+C(p,t, \lambda ))
\ee
where
\be
(1+C(t,p,\lambda )) \delta (p-p')=<p|e^{it(H_0+M)}Te^{-it(H_0+M)}T^*|p'>
\label{Ct}\ee
and $M$ and $T$ should be computed from the recursive relations.

We have proved the following 
\bigskip

{\bf Theorem 6.} For translation invariant Hamiltonians without vacuum
polarization the one particle matrix elements of evolution operator 
are given by the formula 
(\ref{ABc}) where functions $M(p,\lambda )$ and $Z(p,\lambda )$
are solutions of equations (\ref{TW})-(\ref{M}) and (\ref{T*}).
Function $C(t,p,\lambda )$ is defined by (\ref{Ct})
and it vanishes as $t \to \infty$.

\bigskip

\section{Conclusion}

We have obtained in this paper the explicit representations
for the vacuum and one-particle matrix elements of the evolution operator.
By using these representations we have computed the corrections to the 
known results for the large time exponential behaviour
of these matrix elements. This opens the way for 
further investigations of the large time behaviour in 
quantum theory. In particular the problems of quantum decoherence
and decay require the further study by using these methods.

 \bigskip

{\bf Acknowledgments.}  This work is
supported in part by INTAS grant 96-0698,  I.Ya.A. 
 is supported also by RFFI-99-01-00166
and I.V.V. by RFFI-99-01-00105 and by grant for the leading
scientific schools.

\newpage
{\Large{\bf APPENDIX}}

\appendix
\section{ 4-th order for hamiltonian with vacuum polarization}
\setcounter{equation}{0}
For an illustration of the form (\ref{n-or}) of vacuum expactation
value of evolution operator 
we present on Fig.\ref{Fig4-or}
all contributions to the 4-th order of perturbation theory for  
vacuum expactation
value of evolution operator. For simplicity we consider the interaction in
the form 
\be
V=\int (v(p_1,p_2 p_3)(a^*(p_1)+a(p_1))((a^*(p_2)+a(p_2))a^*(p_3)+a(p_3))
dp_1dp_2dp_3
\label{kub}
\ee

On this example  we see once again that only one term 
 from all terms represented on
 Fig.\ref{Fig4-or} has linear dependens on $t$. This term is represented by
bracket-diagram on which all right bracked are on the  right
from the last vertex.

\begin{center}
\parbox{6cm}{$~~~~~~$Diagrams}
\parbox{10cm}{ Corresponding order of application of $\Gamma$-operator,\\
 exponential factors and denominators:}
\end{center}

\bigskip

\begin{center}
\parbox{5cm}{\epsfig{file=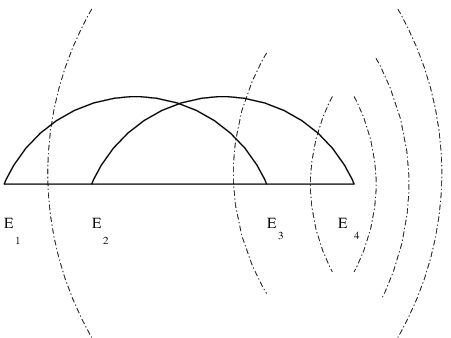,
  angle=0
 }}
\parbox{11cm}{
$T<0|V\Gamma(V\Gamma(V\Gamma(V))))|0>=$\\
$~$\\
$T\int dp...\frac{(\overline v\circ\overline v\circ v\circ v)(p,..)}
{(E_1+E_2+E_3+E_4)(E_2+E_3+E_4)
(E_3+E_4)E_4}$}
\end{center}

\bigskip

\begin{center}
\parbox{5.5cm}{\epsfig{file=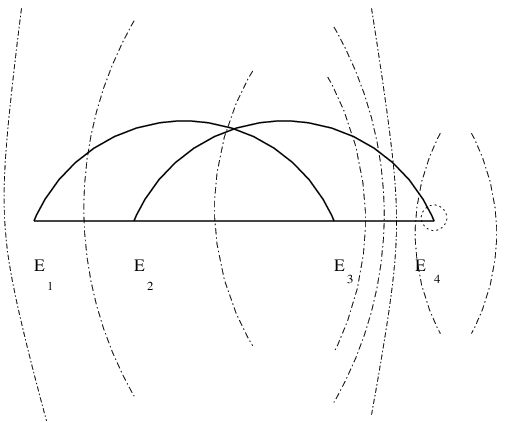,
  angle=0
 }}
 \parbox{11cm}{
$-\int dp...[e^{i(E_1+E_2+E_3)T}-1]\cdot 
\Gamma(v\circ \Gamma(v\circ )\Gamma(v\circ ))\Gamma(v)(p,...)=$\\
$~$\\
$-\int dp...[e^{i(E_1+E_2+E_3)T}-1]
\frac{(\overline v\circ\overline v\circ v\circ v)(p,..)}
{(E_1+E_2+E_3+io)(E_2+E_3+io)(E_3+io)E_4}$}
\end{center}

\bigskip

\begin{center}
\parbox{5.5cm}{\epsfig{file=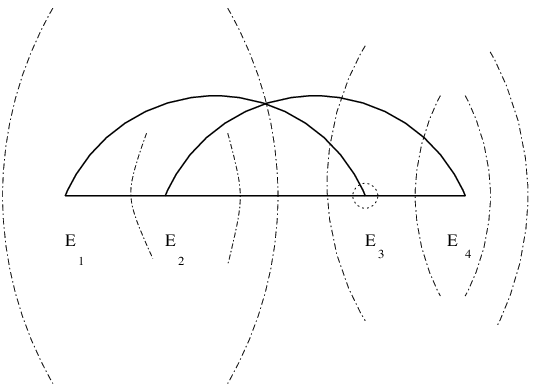,
  angle=0
 }}
 \parbox{11cm}{
$-\int dp...[e^{i(E_1+E_2)T}-1]\cdot\Gamma(\overline v\circ\Gamma(\overline v\circ))
\Gamma(v\circ\Gamma(v))(p,...)$\\
$~$\\
$-\int dp...[e^{i(E_1+E_2)T}-1]\cdot 
\frac{(\overline v\circ\overline v\circ v\circ v)(p,..)}
{(E_1+E_2+io)(E_2+io)
(E_3+E_4+io)E_4}$}
\end{center}

\bigskip

\begin{center}
\parbox{5.5cm}{\epsfig{file=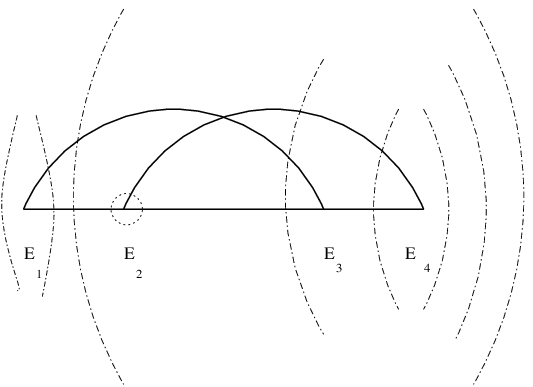,
  angle=0
 }}
 \parbox{11cm}{
$-\int dp...[e^{i(E_1)T}-1]\Gamma(\overline v\circ)\Gamma(\overline v\circ
\Gamma(\overline v\circ\Gamma(v)))(p,..)=$\\
$~$\\
$-\int dp...[e^{i(E_1)T}-1]\cdot 
\frac{(\overline v\circ\overline v\circ v\circ v)(p,..)}
{ (E_1+io)](E_2+E_3+E_4+io)
(E_3+E_4+io)E_4}$}
\end{center}

\bigskip

\begin{center}
\parbox{5.5cm}{\epsfig{file=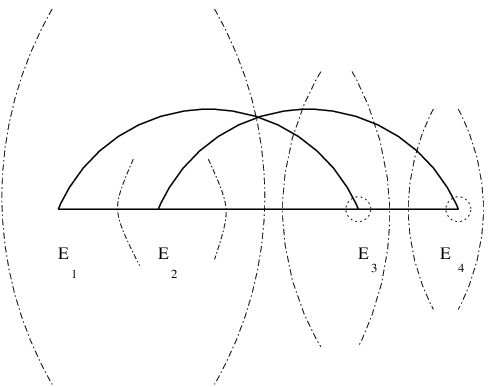,
  angle=0
 }}
 \parbox{11cm}{
$-\int dp...[e^{i(E_1+E_2)T}-1]\cdot\Gamma(\overline v\circ
\Gamma(\overline v\circ))\Gamma(v\circ)\Gamma(v)(p,..)=$\\
$~$\\
$-\int dp...[e^{i(E_1+E_2)T}-1]\cdot
\frac{(\overline v\circ\overline v\circ v\circ v)(p,..)}
{ (E_1+E_2+io)(E_2+io)
(E_3+io)E_4}$}
\end{center}

\bigskip

\begin{center}
\parbox{5.5cm}{\epsfig{file=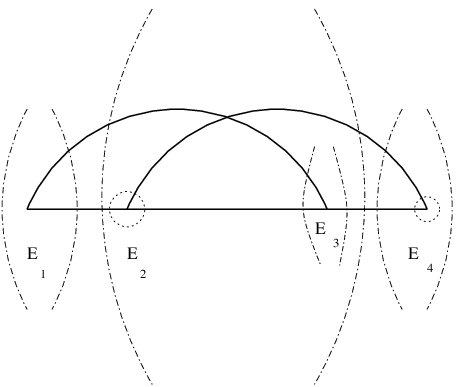,
  angle=0
 }}
 \parbox{11cm}{
$\int dp...[e^{iE_1T}-1]\cdot|
\Gamma(\overline v\circ)\Gamma(\overline v\circ
\Gamma(v\circ))\Gamma(v)(p,..)=$\\
$~$\\
$\int dp...[e^{iE_1T}-1]\cdot
\frac{(\overline v\circ\overline v\circ v\circ v)(p,..)}
{E_1(E_2+E_3+io)
(E_3+io)E_4}$}
\end{center}

\bigskip

\begin{center}
\parbox{5.5cm}{\epsfig{file=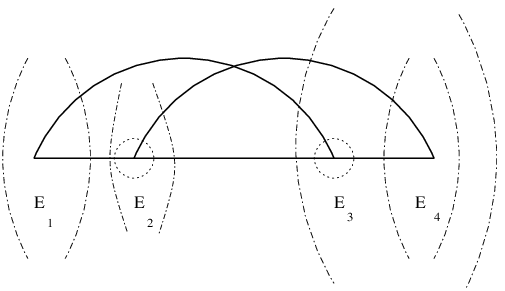,
  angle=0
 }}
 \parbox{11cm}{
$\int dp...[e^{iE_1T}-1]\cdot\Gamma(\overline v\circ)
\Gamma(\overline v\circ)\Gamma(v\circ\Gamma(v))(p,..)$\\
$~$\\
$\int dp...[e^{iE_1T}-1]\cdot
\frac{(\overline v\circ\overline v\circ v\circ v)(p,..)}
{E_1(E_2+io)
(E_3+E_4+io)E_4}$}
\end{center}

\bigskip

\newpage

\begin{center}
\parbox{5.5cm}{\epsfig{file=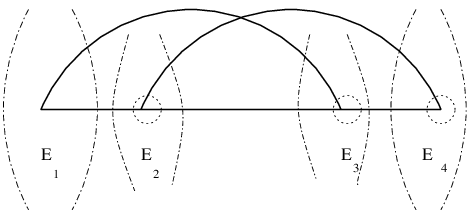,
  angle=0
 }}
 \parbox{11cm}{
$-\int dp...[e^{iE_1T}-1]\cdot
\Gamma(\overline v\circ)\Gamma(\overline v\circ)
\Gamma(v\circ)\Gamma(v)(p,..)$\\
$~$\\
$-\int dp...[e^{iE_1T}-1]\cdot 
\frac{(\overline v\circ\overline v\circ v\circ v)(p,..)}
{E_1(E_2+io)(E_3+io)E_4}$}
\end{center}

\begin{figure}[h]
\caption{4-th order diagrams of vacuum expactation
value of evolution operator for interaction (\ref{kub})}
\label{Fig4-or}
\end{figure}


\section{4-th order of one particle expactation
value of evolution operator for translation invariant interaction }
\setcounter{equation}{0}

Here we prove  the representation (\ref{N7}) - (\ref{N12R})
 for one 
particle matrix elements of evolution operator 
for translation invariant Hamiltonian (\ref{H8}) without 
vacuum polarization
 in the 4-th order of perturbation theory. We obtain 

$$
A_4(p)=A_4^{1PI}(p)+A_4^{1PR}(p),
$$

\be
A_4^{1PI}(p)=(V\Gamma(V\Gamma(V\Gamma(V))))^{1PI}_{1,1}(p),
\label{N12aa}
\ee

\be
A_4^{1PR}(p)=(V\Gamma(V)(V\Gamma^2(V)))^{1PR}_{1,1}(p)
\label{N12Ra}
\ee

 There are two one-particle irreducible (1PI) diagrams
(Fig. \ref{Fig4-orI}) and one
 one-particle reducible (1PI) diagram,
(Fig. \ref{Fig5}). The contribution of the 4-th order 1PI-diagram
is

$${\cal U}^{1PR}_4(t,p,q,k)=(-i\lambda)^4\int^T_0dt_1
\int^{t_1}_0dt_2\int^{t_2}_0dt_3
\int^{t_3}_0dt_4|v(p,q)|^2\cdot
|v(p,k)|^2e^{iE_1(t_1-t_2)}e^{iE_4(t_4-t_3)}=$$

$$
(-i\lambda)^2 \int^t_0 dt_1\int^{t_1}_0dt_2 |v(p,k)|^2
e^{iE_1(t_1-t_2)}{\cal U}_2(t,p,q)$$

Substituting here (\ref{U_2})
we have

\be
{\cal U}^{1PR}_4(t)(p,q,k)={(iT)^2\over2}\,
{\cal A}_2(p,q){\cal A}_2(p,k)
\label{U4}
\ee

$$-iT\,[B_2(p,q) {\cal A}_2(p,k)
+{\cal A}_4^{(1PR)}(p,q,k)]+{\cal B}^{1PR}_4(p,q,k)+
{\cal C}^{1PR}_4(t,p,q,k)
$$
where
\be
{\cal A}_2(p,q)=(\overline v\circ\Gamma(v))(p,q)
\ee

\be
{\cal A}_4^{(1PR)}(p,q,k)=(\overline v\circ\Gamma(v))(p,q) 
 (\overline v\circ\Gamma^2(v))(p,k)
\ee

$$
{\cal B}_4^{(1PR)}(p,q,k)=[(\overline v\circ\Gamma^3(v))(p,k)
(\overline v\circ\Gamma(v))(p,q)+
(\overline v\circ\Gamma^2(v))(p,k)(\overline v\circ\Gamma^2(v))(p,q)]
+
$$

\be
(\overline v\circ\Gamma(\Gamma^2(v)\circ
\overline v)\circ\Gamma(v))(p.q,k)+
(\overline v\circ\Gamma(\Gamma(v)\circ
\overline v)\circ\Gamma^2(v))(p.q,k)
\ee

$$
C_4^{(1PR)}(t,p,q,k)=
-e^{iE_1T}~[(\overline v\circ\Gamma^3(v))(p,k)
(\overline v\circ\Gamma(v))(p,q)+
(\overline v\circ\Gamma^2(v))(p,k)(\overline v\circ\Gamma^2(v))(p,q)]
$$

\be
e^{iE_1t}(\overline v\circ\Gamma(\Gamma^2(v)\circ
\overline v)\circ\Gamma(v))(p.q,k)+
e^{iE_3t}(\overline v\circ\Gamma(\Gamma(v)\circ
\overline v)\circ\Gamma^2(v))(p.q,k),
\ee
here $E_1=\omega (p)-\omega (q)-\omega (p-q)$ and 
$E_3=\omega (p)-\omega (k)-\omega (p-k)$.

(\ref{U4}) can be rewritten in more compact form
 \be
\label{U4c}U^{1PR}_{1,1~(4)}(t,p)={(it)^2\over2}\,
 A^2_2(p)
-it\,[B_2(p)  A_2(p)
+ A_4^{(1PR)}(p)]+ B^{1PR}_4(p)+
C^{1PR}_4(T,p)
\ee
where
\be
 A_2(p)=(V\Gamma(V))_{1,1}(p), ~~~
A_4^{(1PR)}(p)=(V\Gamma(V))_{1,1}(p) 
 (V\Gamma^2(V))_{1,1}(p)
\ee

\be
B_4^{(1PR)}(p)=(V\Gamma^3(V))_{1,1}(p)
(V\Gamma(V))_{1,1}(p)+
(V\Gamma^2(V))_{1,1}(p)(V\Gamma^2(V))_{1,1}(p)+
\ee

$$
( V\Gamma(\Gamma^2(V)V)\Gamma(V))_{1,1}^{1PR}(p)+
(V\Gamma(\Gamma(V)V)\Gamma^2(V))_{1,1}^{1PR}(p)
$$

$$
C_4^{(1PR)}(t,p)=-(e^{iH_0T}V\Gamma^3(V))_{1,1}(p)
(V\Gamma(V))_{1,1}(p)-
(e^{iH_0t}V\Gamma^2(V))_{1,1}(p)(V\Gamma^2(V))_{1,1}(p)
+$$

$$
( e^{iH_0t}V\Gamma(\Gamma^2(V)V)\Gamma(V))_{1,1}^{1PR}(p)+
(V\Gamma(\Gamma(V)V)\Gamma^2(V)E^{-iH_0t})_{1,1}^{1PR}(p)
$$

To 4-th order of 1-particle average  of wave operator
give also contributions the 1PI diagrams, see Fig.\ref{Fig4-orI}. 
The leading terms
corresponding to these diagrams are of order t.
We have 
\be
 U^{1PI}_4(t,p)=it(V\Gamma(V\Gamma(V  \Gamma (V))))_{1,1}^{1PI}(p)
\ee
Above consideration proves that up to  the 
$t^2$-order  the formula (\ref{N7})
is true.

\newpage
\end{document}